\documentclass[linenumbers,trackchanges,twocolumn]{aastex62}

\newcommand{\revise}[1]{{\color{black}#1}}
\newcommand{\revisev}[1]{{\color{black}#1}}

\usepackage{comment}

\usepackage{amsmath}
\usepackage{hyperref}
\usepackage{natbib}
\usepackage{color}
\usepackage{graphicx}
\usepackage{amssymb}
\usepackage{bm}
\usepackage{tabularx}
\usepackage{yfonts}

\newcommand{\sao}{\affiliation{Smithsonian Astrophysical Observatory, Cambridge, MA, USA}}

\newcommand{\uofa}{\affiliation{University of Arizona, Tucson, AZ, USA}}
\newcommand{\unh}{\affiliation{University of New Hampshire, Durham, NH, USA}}
\newcommand{\ucl}{\affiliation{Mullard Space Science Laboratory, University College London, Holmbury St. Mary, Dorking, RH5 6NT, UK}}
\newcommand{\uchicago}{\affiliation{University of Chicago, Chicago, IL, USA}}

\begin{document}
\nolinenumbers

\title{The Effects of Non-Equilibrium Velocity Distributions on Alfv\'en Ion-Cyclotron Waves in the Solar Wind}

\author[0000-0001-9717-8718]{Jada Walters}\uofa
\author[0000-0001-6038-1923]{Kristopher G. Klein}\uofa
\author[0000-0003-1945-8460]{Emily Lichko}\uofa\uchicago
\author[0000-0002-7728-0085]{Michael L. Stevens}\sao
\author[0000-0002-0497-1096]{D. Verscharen}\ucl
\author[0000-0003-4177-3328]{B. D. G. Chandran}\unh

\begin{abstract}
\nolinenumbers

In this work, we investigate how the complex structure found in solar wind proton velocity distribution functions (VDFs), rather than the commonly assumed two-component bi-Maxwellian structure, affects the onset and evolution of parallel-propagating microinstabilities. 
We use the Arbitrary Linear Plasma Solver (\texttt{ALPS}), a numerical dispersion solver, to find the real frequencies and growth/damping rates of the Alfvén modes calculated for proton VDFs extracted from Wind spacecraft observations of the solar wind. 
We compare this wave behavior to that obtained by applying the same procedure to core-and-beam bi-Maxwellian fits of the Wind proton VDFs. 
We find several significant differences in the plasma waves obtained for the extracted data and bi-Maxwellian fits, including a strong dependence of the growth/damping rate on the shape of the VDF. 
By application of the quasilinear diffusion operator to these VDFs, we pinpoint resonantly interacting regions in velocity space where differences in VDF structure significantly affect the wave growth and damping rates. This demonstration of the sensitive dependence of Alfv\'en mode behavior on VDF structure may explain why the Alfv\'en ion-cyclotron instability thresholds predicted by linear theory for bi-Maxwellian models of solar wind proton background VDFs do not entirely constrain spacecraft observations of solar wind proton VDFs, such as those made by the Wind spacecraft.
\end{abstract}


\section{Introduction}
\label{sec:intro}

\begin{figure*}[t]
    \includegraphics[width=\textwidth]{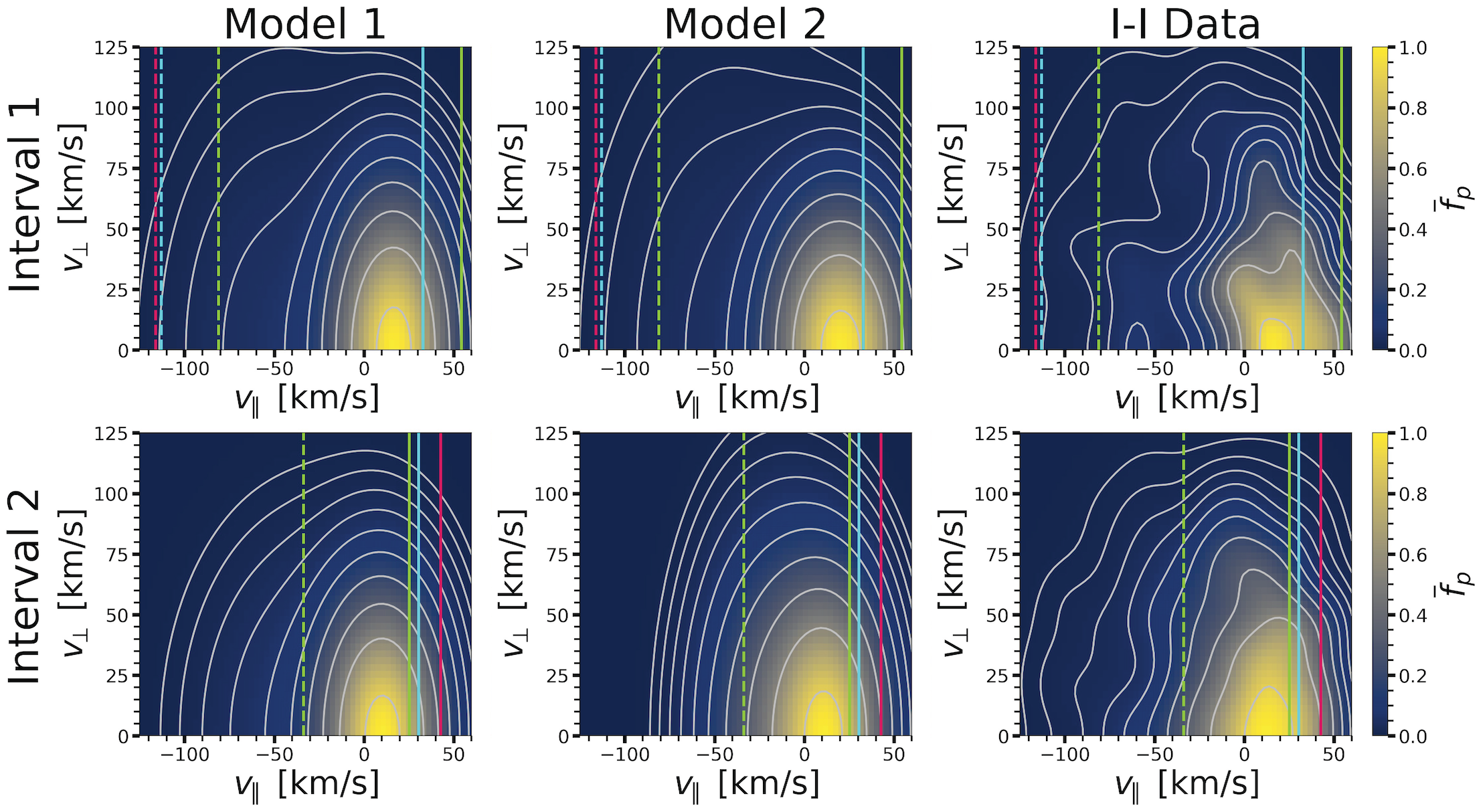}
    \caption{\normalsize{Bi-Maxwellian core-and-beam models (left and center columns) and I-I data (right column) for Intervals 1 (top row) \& 2 (bottom). Vertical lines correspond to the resonant velocity at which an instability terminates for the I-I data (red), first bi-Maxwellian model (blue), and second bi-Maxwellian model (green). Dashed and solid lines indicate resonant velocities associated with the forward and backward Alfv\'en modes respectively. The velocity range  is trimmed in this figure to highlight the structure of the VDF, with the actual velocity ranges used in the \texttt{ALPS} calculation spanning  $v_\parallel \in [-207,205]$ km/s and and $v_\perp \in [0,378]$ km/s for Interval 1 and $v_\parallel \in [-204,202]$ km/s and $v_\perp \in [0,356]$ km/s for Interval 2. While the color scale is linear, gray contours show the shape of $\bar{f}_p$ across 10 logarithmically-spaced levels between 0.01 and 0.9}}
    \label{fig:intVDF}
\end{figure*}

Space and astrophysical plasmas are commonly weakly collisional. With infrequent collisions, the velocity distribution functions (VDFs) of these plasmas can significantly differ from the entropically-favored Maxwellian shape  \citep{Vasyliunas:1968, Gosling:1981, Lui:1981, Marsch:1982, Armstrong:1983, Lui:1983, Williams:1988}. These departures from equilibrium can act as sources of free energy that drive instabilities. Instabilities play an important role in both the small- and large-scale plasma behavior, e.g. by pitch-angle scattering of particles acting as an effective viscosity \citep{Kunz:2011,Riquelme:2018,Arzamasskiy:2022} or by serving as channels to transport energy in space; see the review by \cite{Matteini:2012}. In the weakly collisional plasma that comprises the solar wind a variety of non-equilibrium features develop, including temperature anisotropy \citep{Kasper:2002}, temperature disequilibrium between species \citep{Neugebauer:1976}, and beams \citep{Alterman:2018, Pilipp:1987}.

To better describe the non-equilibrium features in solar wind VDFs, it is common to fit observations to simplified model distributions such as bi-Maxwellian \citep{Marsch:2006} or kappa \citep{Summers:1994} distributions. 
Bi-Maxwellian models have been extensively used to determine instability thresholds as a function of proton temperature anisotropy ($T_{\perp, p} / T_{\parallel,p}$ where perpendicular/parallel are defined with respect to the mean magnetic field $B$) and the parallel proton plasma beta ($\beta_{\parallel,p} = 8\pi n_p k T_{\parallel,p} / B^2$ where $n_j$ is the density of a plasma component $j$). Assuming a bi-Maxwellian shape of the background VDFs and a homogeneous background, different kinds of wave instabilities are commonly expected to be driven beyond these instability thresholds. The thresholds can account for either a single free energy source \citep[e.g.][]{Hellinger:2006, Bale:2009} or multiple sources \citep[e.g.][]{Matteini:2013,Chen:2016,Klein:2018}; reviews of these thresholds are provided by \cite{Gary:1993, Verscharen:2019}. 

A well-known open question in heliophysics\revise{, and the focus of this paper,} concerns the Alfv\'en ion-cyclotron (AIC) instability, which occurs when the proton temperature is larger perpendicular to the mean magnetic field direction than parallel, $T_{\perp,p} > T_{\parallel,p}$ \citep{Kennel:1967, Davidson:1975}. The predictions of linear theory for the AIC instability threshold do not fully constrain the observations of solar wind proton VDFs by the Wind spacecraft when calculated for bi-Maxwellian proton background VDFs \citep{Hellinger:2006,Bale:2009}. Several explanations have been proposed for the failure of this approach to describe the observations, including inefficient energy extraction \citep{Shoji:2009} or the impact of other plasma populations such as $\alpha$-particles \citep{Maruca:2012}. \cite{Isenberg:2013} suggest that the assumption of bi-Maxwellian proton distributions limits the applicability of the thresholds predicted by bi-Maxwellian linear theory. Our investigation focuses on understanding how the structure of realistic solar wind VDFs affects the onset and evolution of the AIC instability in the solar wind compared to the bi-Maxwellian approximation.

The structure of a plasma VDF has physical consequences in terms of growth or damping of plasma waves, which are important carriers of energy in collisionless plasmas such as the solar wind. We are therefore interested in understanding how the properties of waves differ between those produced from core-and-beam bi-Maxwellian models of spacecraft observations of solar wind proton VDFs and the spacecraft data itself.

Linear plasma solvers (such as \texttt{WHAMP}, \texttt{NHDS}, or \texttt{PLUME}) designed to study the plasma waves associated with a given background VDF typically use a bi-Maxwellian model and solve the hot-plasma dispersion relation associated with this simplified description of the distribution \citep{Roennmark:1982, Quataert_1998, Klein:2015, Verscharen:2018a}. 
In our investigation of how deviations from an idealized bi-Maxwellian distribution affect the linear properties of the AIC instability, we use \texttt{ALPS} (the Arbitrary Linear Plasma Solver), a code that solves the hot-plasma dispersion relation by direct numerical integration of  arbitrary background distributions for the plasma; a detailed overview of \texttt{ALPS} is provided by \cite{Verscharen:2018}.

In section~\ref{sec:methods} we describe the VDFs used in this work and outline the numerical tools employed to calculate the linear plasma response. 
We present the plasma behavior of the Alfvén modes associated with these VDFs in section~\ref{sec:results}, along with a more detailed analysis of the connection between VDF structure and the stability of the Alfvén modes through calculation of the quasilinear diffusion operator. 
We summarize our results in section~\ref{sec:discussion}.

\section{Methodology}
\label{sec:methods}

\revise{\subsection{VDF Construction}}
\label{ssec:vdfconstruction}

The VDFs used in this paper are based on two representative distributions observed by the Solar Wind Experiment (SWE) aboard the Wind spacecraft \citep{Ogilvie:1995}. The Wind spacecraft rotates with a period of $\approx$ 3 seconds and charge flux distributions, as a function of energy and incidence angle, are acquired over $\approx$ 93 seconds. \revisev{This rotation allows the distribution function to be resolved over the full field of view of $4\pi$ steradians.} The distributions used in this work are based on \revise{observations made over one acquisition period} on 2009 December 19 from 02:49:41 UT and 2008 August 14 from 04:31:43.69 UT for Interval 1 and Interval 2 respectively. The average magnetic field during each period is measured by the Magnetic Field Investigation (MFI) \citep{Lepping:1995} and the average magnitude of the magnetic field over this 93 second interval is 5.044 nT for Interval 1 and 6.065 nT for Interval 2. Based on a nonlinear bi-Maxwellian fit to the SWE proton measurements \citep{Kasper:2002}, we find $\beta_{\parallel,p} =0.153$, $T_{\perp,p}/T_{\parallel,p} =2.41$ (Interval 1) and  $\beta_{\parallel,p} =0.148$, $T_{\perp,p}/T_{\parallel,p} =2.80$ (Interval 2). Both intervals are therefore above the $\gamma/\Omega_p=10^{-2}$ marginal stability threshold (where $\gamma$ is the growth/damping rate, $\Omega_p=q_p B/m_p$ is the proton cyclotron frequency, $q_p$ is the proton charge, and $m_p$ is the proton mass) for the AIC instability if the distribution function were modeled as a bi-Maxwellian, but below the same threshold for the mirror instability \citep{Verscharen:2016}. We choose intervals in this region of $\beta_\parallel$, $T_{\perp,p}/T_{\parallel,p}$ space to allow us to characterize the role of non-equilibrium VDF structure on the stability of solar wind plasma that should be unstable to the AIC instability under the bi-Maxwellian assumption. \revisev{These intervals exhibit a prominent proton beam population; such intervals have historically been treated with core-and-beam bi-Maxwellian fits. The $\alpha$-particle density is below the detection threshold for the Wind non-linear analysis for Interval 1 and around $n_\alpha/n_p \sim 0.007$ for Interval 2. These low $\alpha$-particle densities do not contribute significantly to the linear plasma response, and thus we will not consider the $\alpha$-particles in this analysis.}

The Wind SWE Faraday cup has a baseline resolution of 31 bins logarithmically distributed across voltages from 150 to 8000 V with energy resolution of $\Delta E / E \approx 0.065$ \citep{Ogilvie:1995} We further generalize the bi-Maxwellian VDF model typically extracted from these measurements, provided by \cite{Ogilvie:2021}, to better describe the Wind SWE Faraday Cup observations. The Faraday Cup response function is not uniquely or analytically invertible, so a numerical approach is required in order to recover details of the VDF that are not well-described by the bi-Maxwellian model. We choose to characterize the VDFs more completely by following a two-step approach: first, we capture the dominant features with a double bi-Maxwellian regression; second, we discretize that model and apply random gyrotropic perturbations to improve the match with the data. The resulting discrete gyrotropic models are suitable for analysis with \texttt{ALPS}.

We \revise{first} choose a double bi-Maxwellian VDF model to capture both the core and secondary beam proton components that are evident in the data. The Faraday Cup data is natively expressed as ion charge fluxes measured as a function of voltage and spacecraft spin angle \citep{Kasper:2021}. Following the protocol used in several previous works, \revise{\citep{Alterman:2018, Chen:2016, Jiansen:2018, Wicks:2016, Gary:2016}}, we fit the measured instrument response to the modeled instrument response for two bi-Maxwellian distributions. \revise{We superpose two instances of the Faraday Cup response to a single bi-Maxwellian, where the analytic form for the latter is described by \cite{Kasper:2006} and the sensitive area as a function of angle is derived for each datum by interpolating on the effective area table provided in the metadata of \cite{Kasper:2021}. We then perform regressions to the data, with gyrotropic constraints applied to the bi-Maxwellian symmetry axes and relative drifts.} The resultant best-fit parameters obtained using this method are hereafter referred to as “model 1,” as they are labeled in Table \hyperref[tab:params]{1a}.

To explore non-Maxwellian variations on the fits, we first discretize each model 1 fit to a Cartesian ($v_\perp$, $v_\parallel$) grid with 2 km/s x 2 km/s resolution and then recompute the Faraday Cup response numerically. Calculating the Faraday Cup response requires one to integrate the differential charge flux upon the sensor, multiplied by the effective sensitive area, over an appropriate domain. In this case, we take the differential charge flux associated with any point in phase space from the gridded model and retrieve the effective area from the table provided in \cite{Kasper:2021}. \revise{The integration domain for each datum is the conical frustum in phase space defined by the cup’s full field of view and the bounds of the particular energy window. The integrations are performed using the IDL (Interactive Data Language) “int\_3d” utility with twenty-point Gaussian quadrature.}

\revise{For both spectra, we calculate the non-reduced chi-square statistic by comparing the numerical model to the data and then employ a Monte Carlo strategy. At each iteration, (1) a pseudorandom perturbation is superposed at each point in the discrete VDF model, (2) the Faraday Cup response to the perturbation is calculated, (3) the chi-square is recalculated for the perturbed model, and (4) the perturbation is accepted or rejected on the basis of whether the chi-square was reduced. Gaussian perturbations are generated in an ad hoc manner: the generator selects randomized magnitudes between -0.1 and 0.1 of the local VDF value and half-widths between 1 and 10 grid cells. To select phase space locations, the generator uses the VDF as a probability distribution function, with an ad hoc flattening and artificial enhancement out to six thermal widths. This is found to strongly favor low-energy solutions while still providing a clear and reliable signal at the tails.} We carry out the Monte Carlo procedure until the chi-square statistic converges to 1 part in 10,000. The resulting modified spectrum models are hereafter referred to as the “inverted-interpolated data” or "I-I data" as they are labeled in Fig. \ref{fig:intVDF}.

\begin{figure*}[t]
    \includegraphics[width=\textwidth]{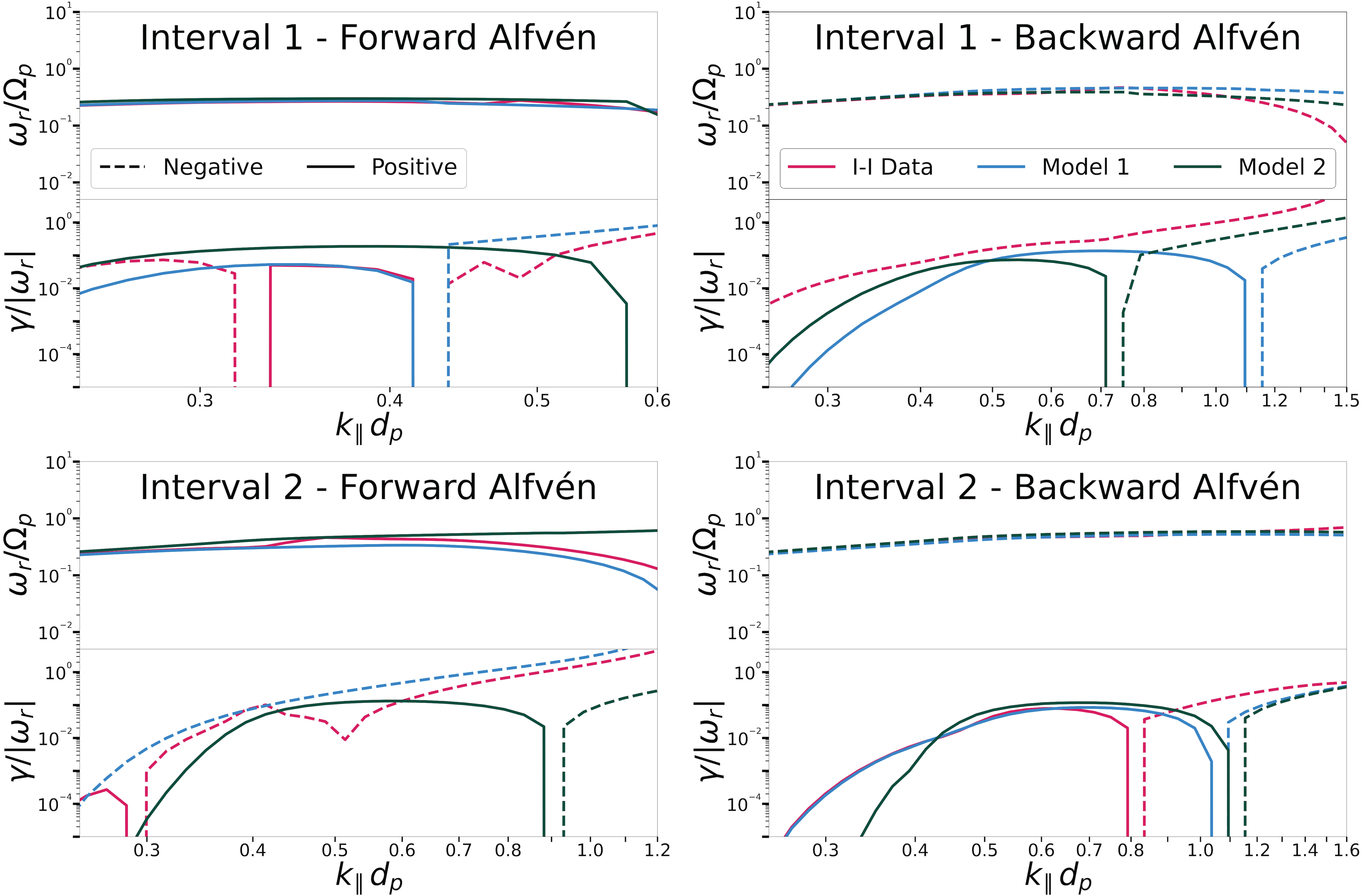}
    \caption{\normalsize{\texttt{ALPS} dispersion solutions for real and imaginary frequencies $\omega_r$ and $\gamma$. The forward (left column) and backward (right) Alfv\'en solutions are presented for Interval 1 (top panels) and 2 (bottom). The $\omega_r$ solutions for the first (blue) and second (green) bi-Maxwellian models are qualitatively similar to the I-I data (red) solution, but the $\gamma$ behavior differs significantly. For instance, the bi-Maxwellian models are unstable for the backward Alfv\'en solution for Interval 1 while the I-I data yields a stable solution. The forward Alfv\'en solution in Interval 2 is unstable in model 2 while model 1 and I-I data are stable. \revise{The \texttt{ALPS} scan is performed for $k_\parallel d_p$ between 0.01 and 10, and we choose the limited range presented here to constrain the scan to the region where the resonant velocity is within the limits of the input distribution and $\gamma / \omega_r < 1 / e$.}}}
    \label{fig:ALPS22}
\end{figure*}

 We also use the Levenberg-Marquardt method to construct core-and-beam bi-Maxwellian fits to the I-I data. These best-fit models are hereafter referred to as “model 2,” as they are labeled in Table \hyperref[tab:params]{1a}. As this work aims to understand the impacts of VDF structure on microinstabilities, we include both of these independently-constructed models to determine whether small differences between bi-Maxwellian models also yield different linear plasma response. The fit parameters for both model 1 and model 2 are shown in Table \hyperref[tab:params]{1a}.
 
 Figure~\ref{fig:intVDF} shows the I-I data and two core-and-beam bi-Maxwellian models for both intervals, where $\bar{f}_p$ is a re-scaling of $f_p$ to a maximum value of 1. The contours highlight how a two-component bi-Maxwellian fit is unable to capture the complex structure of the I-I data, at both small and large $v_\parallel$.
 \begin{figure*}[t]
    \includegraphics[width=\textwidth]{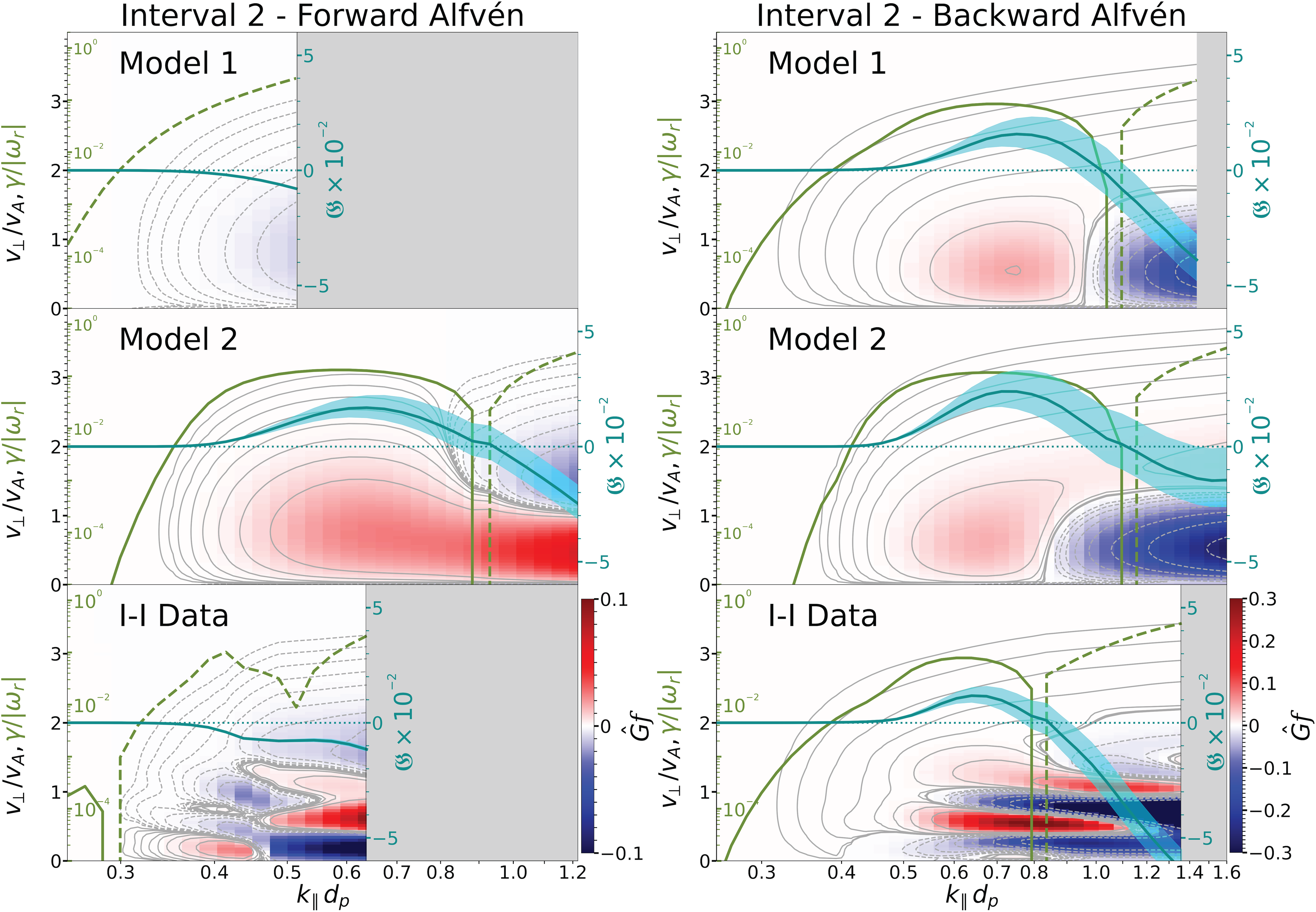}
    \caption{\normalsize{Diffusion operator $\hat{G}f$ (Eqn.~\ref{eq:Ghat}) as a function of $k_\parallel$ and $v_\perp$ indicates regions where energy is transferred from the distribution to the fields (red) or vice versa (blue). Colorbars at bottom row apply to entire column. $\hat{G}f$ is dimensionless but proportional to $(n_j/n_0) / (v/v_A)^4$ where $n_0 \equiv \int d^3 f_j$. $\hat{G}$ calculations are shown for Interval 2 forward Alfv\'en mode (left) and backward Alfv\'en mode (right) for the first bi-Maxwellian model (top), second bi-Maxwellian model (middle), and I-I data (bottom).  $\textfrak{G}$ is plotted against the right-hand y-axis and given by the cyan line with error bounds as described in the text, with the associated growth (solid green line) or damping (dashed) rates plotted against the left-hand y-axis. $\textfrak{G}$ is dimensionless but proportional to $(n_j/n_0) / (v/v_A)^7$. Where $\textfrak{G}$ is positive, we find a corresponding instability (positive $\gamma$). For the forward Alfvén mode, gray contours are 8 logarithmically-spaced levels between $\lvert5e-5\rvert$ and $\lvert.01\rvert$ for both positive (solid) and negative (dashed) values. For the backward Alfvén mode, contours are 8 logarithmically-spaced levels between -0.25 and -1e-2 and 8 logarithmically-spaced levels between 1e-5 and 0.05.}}
    \label{fig:diffusion1}
\end{figure*}
\revise{\subsection{Calculation of the Dispersion Relation}}
\label{ssec:discalc}

We employ \texttt{ALPS} to calculate the linear kinetic plasma behavior of the bi-Maxwellian model and I-I data VDFs. \revise{For full code details, see \cite{Verscharen:2018}.} \texttt{ALPS} seeks non-trivial solutions to the wave equation by \revise{numerical} calculation of the plasma susceptibility of each component species. \revise{Unlike} dispersion solvers that simplify this process by modeling the distributions as bi-Maxwellians (e.g. \texttt{PLUME}, \texttt{NHDS}) or kappa distributions (e.g. \texttt{DSHARK} \citep{Astfalk:2015}), \texttt{ALPS} \revise{numerically} integrates the susceptibility directly from a given gyrotropic background distribution $f_j(v_\perp,v_\parallel)$. \revise{By numerically integrating the phase space density to solve the plasma wave equation rather than using the properties of special functions such as Maxwellians to find closed-form solutions of the dispersion relation, \texttt{ALPS}} allows \revise{us to analyze} the I-I Wind data directly, without making the simplifying assumption that the VDF structure follows a closed analytical form.

\begin{figure*}[t]
    \includegraphics[width=\textwidth]{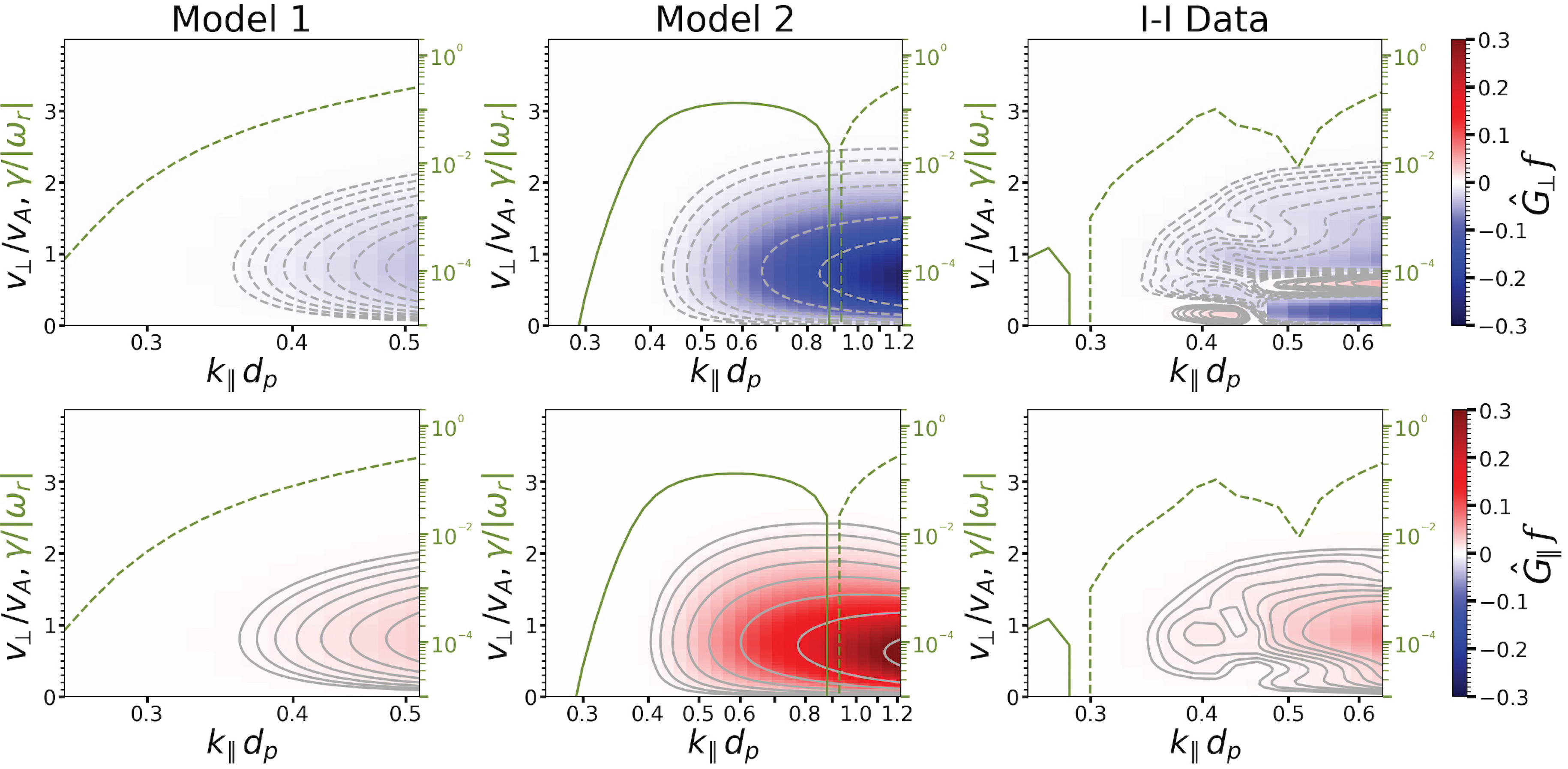}
    \caption{\normalsize{Perpendicular (top row) and parallel (right) diffusion operators, Eqn.~\ref{eq:Gperp} and Eqn.~\ref{eq:Gparallel}, for the bi-Maxwellian models 1 and 2 (left and middle columns) and I-I data (right) for the Interval 2 forward Alfv\'en solution. The color scheme follows that presented in Figure~\ref{fig:diffusion1}. The models have qualitatively similar structures, but exhibit different $\hat{G}_\perp f$ and $\hat{G}_\parallel f$. This can be compared against the complexity seen in the I-I data, where the same broad structures are not observed. For model 1 and the I-I data, gray contours are 8 logarithmically-spaced levels between $\lvert5e-4\rvert$ and $\lvert0.03\rvert$ for both positive (solid) and negative (dashed) values. For model 2, gray contours are 8 logarithmically-spaced levels between $\lvert5e-3\rvert$ and $\lvert0.3\rvert$ for positive and negative values.}} 
    \label{fig:diffusionperppar}
\end{figure*}

\revise{To numerically solve the hot plasma dispersion relation, laid out by \cite{Stix:1992}, \texttt{ALPS} uses the distributions $f_j$ for each species j defined on a discrete grid in perpendicular and parallel momentum space, with minimum and maximum values $P_{min, j}$ and $P_{max, j}$ and resolution defined by the number of steps $N_\perp$ or $N_\parallel$ between $P_{min, j}$ and $P_{max, j}$ in each direction. The calculation of the susceptibility $\chi_j$ is given in equation 2.9 of \cite{Verscharen:2018}. A user defined parameter $J_{\textrm{max}}$ determines the order of resonances $n_{\textrm{max}}$ to include in the calculation of $\chi_j$ such that the maximum amplitude of the Bessel function $J_{n_{\textrm{max}}+1}<J_{\textrm{max}}$. Modes that resonantly interact with portions of the distribution outside the resolved numeric grid are unaffected by the particular shape of $f_j$ and have been shown to  match fluid solutions. 

For damped modes, the wave equation requires an analytic continuation, for which we apply the \emph{hybrid-analytic continuation} scheme described in section 3.2 of \cite{Verscharen:2018}. The Landau contour integration in this scheme is computed by decomposing the integral into a compound function, where numerical integration is used when possible and the residue component of the function is computed using an algebraic function fit to $f_j$ that can be selected by the user. Additional numeric methods are needed to integrate the poles; these are described in section 3.1 of \cite{Verscharen:2018}. For these integrations, $t_{\textrm{lim}}$ and $M_I$ are user defined parameters determining the regions over which these numerical procedures are implemented.}

For our calculation of the dispersion relation using \texttt{ALPS}, we use the following parameters. 
For all cases, $J_{max} = 10^{-50}$,  $M_I = 5$, $M_p = 100$, $T_{lim} = 0.01$, and $P_{min,\perp j} = 0 m_pv_A$.
For Interval 1, the momentum space resolution is \revise{$N_\perp = 189$} and \revise{$N_\parallel = 206$}.  For Interval 2, the momentum space resolution is \revise{$N_\perp = 178$} and \revise{$N_\parallel = 203$}. The proton and electron momentum space ranges are presented in Table \hyperref[tab:params]{1b}. The electron VDFs used for the dispersion calculation in \texttt{ALPS} were constructed using the observed electron temperature from Wind SWE (13 eV for Interval 1 and 12.5 eV for Interval 2), with relative drift and density chosen to preserve current-free and quasi-neutral conditions.

\begin{table*}[ht]
    \centering
        \begin{tabular}{c}
        \includegraphics[width=.94\textwidth]{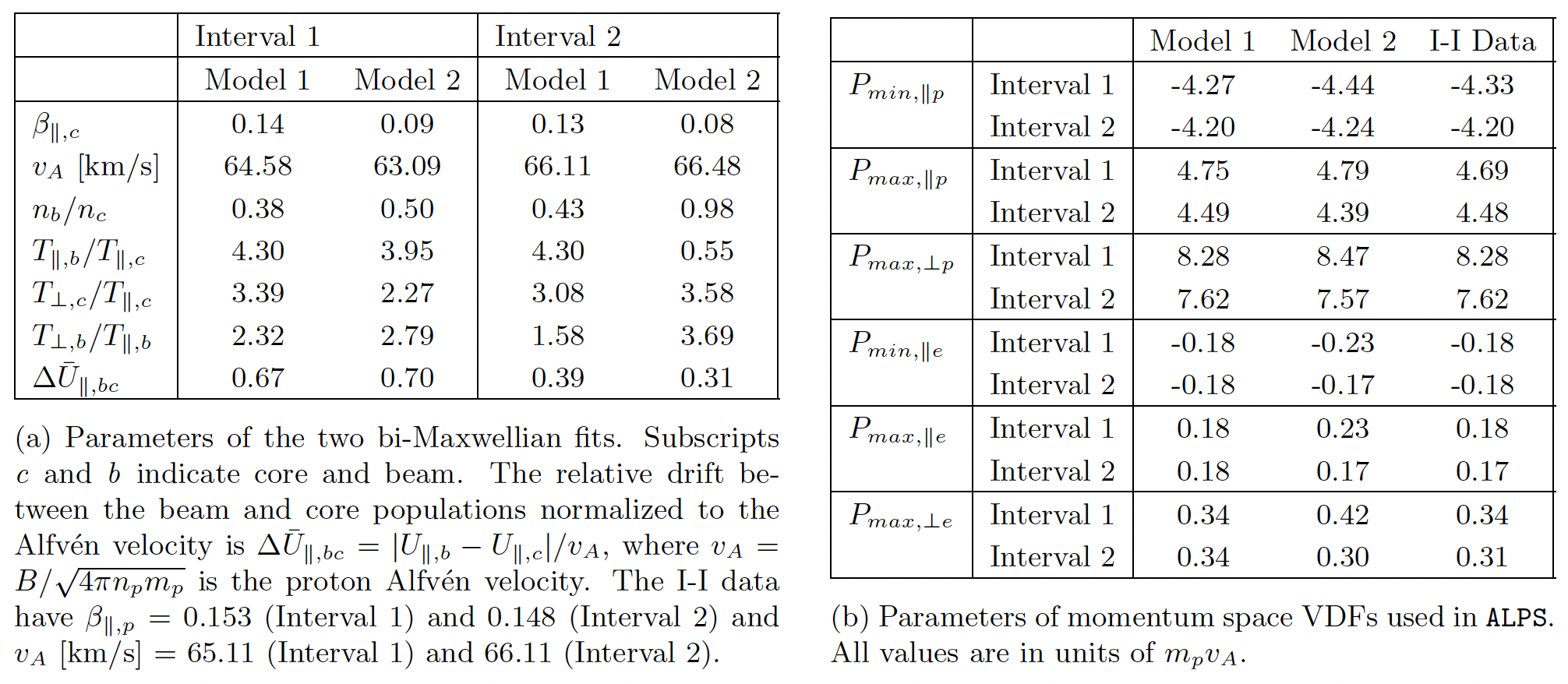}
    \end{tabular}
    \caption{\normalsize{Plasma and \texttt{ALPS} parameters used in this work.}}
    \label{tab:params}
\end{table*}

\section{Results}
\label{sec:results}

\subsection{Dispersion Relation}
\label{ssec:dispersion}

 Figure~\ref{fig:ALPS22} presents the real and imaginary components of the linear mode frequencies, $\omega_r$ and $\gamma$ respectively, for the forward and backward Alfv\'en modes as a function of the normalized parallel wavenumber, $k_\parallel d_p$ (where $d_p = v_{A}/\lvert\Omega_p\rvert$ is the proton inertial length). We calculate the dispersion relation for the parallel-propagating Alfv\'en modes with normalized perpendicular wavenumber $k_\perp d_p = 0.001$ and $k_\parallel d_p$ between 0.01 and 10. The $k_\parallel d_p$ ranges shown in Fig.~\ref{fig:ALPS22} start at $k_\parallel d_p = 0.25$ as this is the lower limit where the resonant velocity $v_{res} = \left({\omega_r - n\Omega_p}\right) / {k_\parallel}$ (where $n$ is an integer) exceeds the $v_\parallel$ bounds of the input distributions. For the upper limit of $k_\parallel$, we \revise{ require that $\gamma / \omega_r < 1 / e$} for consistency with our later calculations of the quasilinear diffusion operator. While the calculated real frequency $\omega_r$ agrees for the data and bi-Maxwellian models for all modes (with slight differences emerging as the waves approach small scales $k_\parallel d_p \gg 1$), there is significant disagreement in $\gamma$. The backward Alfv\'en mode for Interval 1 shows unstable behavior --- a positive $\gamma$ --- for both bi-Maxwellian models, but the solution for the I-I data is stable for all $k_\parallel d_p$. The forward Alfv\'en mode for Interval 2 shows stable behavior in the data and model 1, but model 2 develops an instability. 

Even in the cases for which all models predict an instability, the extent of the unstable wavenumber range in $k_\parallel d_p$ differs. The forward Alfv\'en mode in Interval 1 exhibits only a small region of instability from $k_\parallel d_p =$ 0.33 to 0.41 for the I-I data, compared to both bi-Maxwellian model instabilities which range from $k_\parallel d_p =$ 0.25 to 0.41  (model 1) and $k_\parallel d_p =$ 0.25 to 0.57 (model 2). The backward Alfv\'en mode in Interval 2 has instabilities that span similar $k_\parallel d_p$ ranges for the bi-Maxwellian models, with $k_\parallel d_p =$ 0.25 to 1.04 (model 1) and $k_\parallel d_p =$ 0.33 to 1.09 (model 2), but the instability in the I-I data spans only $k_\parallel d_p =$  0.25 to 0.79, terminating at a smaller $k_\parallel d_p$ than either of the models.

To tie the differences in unstable behavior back to the structure of the input VDFs, we consider the resonant velocities associated with the wavenumber at which the instabilities terminate. These are indicated in Figure~\ref{fig:intVDF} by the vertical colored lines, with any $v_{res} < 0$ corresponding to the forward Alfv\'en modes and $v_{res} > 0$ corresponding to the backward Alfv\'en modes. In Interval 1, the forward Alfv\'en mode in the I-I data interacts with a region further from the core than either of the models, by which point the model VDF contours are quite smooth. The backward Alfv\'en modes for both models resonant with a region of smooth $f_p$ structure for $v_\parallel > 0$ and large temperature anisotropy. The I-I data exhibit a more uneven structure and steeper drop-off of $f_p$ in this same region, and no instability in the backward Alfv\'en wave is found. 

For the Interval 2 forward Alfv\'en mode, we find an instability in model 2, but not model 1, with the width of the core population being the primary difference in VDF structure between the models. While the \texttt{ALPS} results for the I-I data exhibit an instability at -174 km/s, this is very close to the bounds of the input VDF and is likely an artifact of the \texttt{ALPS} scan moving into the resolved VDF range, so we consider this to be an overall stable mode. In the backward Alfv\'en mode, all three resonant velocities shown in Fig. \ref{fig:intVDF} are close to the core population, with the I-I data instability terminating at a larger $v_{res}$ than either model, where the I-I data VDF has a much more varied structure than either model.

\subsection{Diffusion Operator}
\label{ssec:diffusion}
To better understand what drives the differences in the linear plasma response between the I-I data and bi-Maxwellian models, we examine how the differences in the structure of the input distributions affect the sign and magnitude of $\gamma$. We use Eqn.~\ref{eq:ALPS}, an analytical expression for the quasilinear growth rate \revise{$\gamma_j$ of a species $j$} at a particular wavenumber. 
\begin{align}
    \frac{\revise{\gamma_j}}{\lvert {\omega_{r} \rvert}}  = \sum_j \frac{\pi}{8n_j} &\left| \frac{\omega_{r}}{k_\parallel}\right| \left(\frac{\omega_{pj}}{\omega_{r}}\right)^2 \sum_{n=-\infty}^{\infty} \int_0^{\infty} dv_\perp v_\perp^2  \label{eq:ALPS} \\ 
    &\int_{-\infty}^{\infty} dv_\parallel \delta\left(v_\parallel - \frac{\omega\revise{_{r}} - n\Omega_j}{k_\parallel}\right) \frac{\lvert \phi_{n}^2 \rvert \hat{G} f_j }{W}   \nonumber
\end{align}
where $\omega_{pj} = \sqrt{4\pi n_j Z^2 q_j^2 / m_j}$ is the plasma frequency (with $Z$ the charge state) and the diffusion operator, $\hat{G}$, is given by 
\begin{align} 
\hat{G} &\equiv  \hat{G}_{\perp} + \hat{G}_{\parallel},
\label{eq:Ghat}\\
\hat{G}_{\perp} &\equiv \left( 1 - \frac{k_\parallel v_\parallel}{\omega\revise{_{r}}}\right)\frac{\partial}{\partial v_\perp}, \quad \mathrm{and}
\label{eq:Gperp}\\
\hat{G}_{\parallel} &\equiv \frac{k_\parallel v_\perp}{\omega_{r}}\frac{\partial}{\partial v_\parallel} \qquad . 
\label{eq:Gparallel}
\end{align} 
A full derivation of Eqn.~\ref{eq:ALPS} from the Vlasov-Maxwell equations is provided by \cite{Kennel:1967}. Eqn.~\ref{eq:ALPS} is valid when $\revise{\gamma_j} \ll \omega_{r}$. The explicit dependence of \revise{$\gamma_j$} on the distribution function and the resonant velocity links VDF structure to the resultant stability or instability of a given plasma mode. The only component of Eqn.~\ref{eq:ALPS} that is not positive definite is $\hat{G}f_j$, so this term determines whether a species contributes at the given wavenumber range through a positive or negative $\gamma_j$, directly affecting the stability of a given mode based on the structure of $f_j$ (specifically on the $v_\perp$ and $v_\parallel$ gradients of $f_j$ seen in Eqns.~\ref{eq:Gperp} and ~\ref{eq:Gparallel}). The electrons have a minor contribution and the intervals chosen have low \revisev{$\alpha$-particle} density, so we consider only proton growth rate, $\gamma_p$ in our calculation. We neglect positive definite terms that do not impact the sign of the energy transfer, including the pre-factors before the sum over $n$ and the wave energy density \revise{$W$}. Evaluation of Eqn.~\ref{eq:ALPS} using the \texttt{ALPS} Alfv\'en solutions for the three VDFs allows us to better understand what aspects of the VDF structure lead to the strong variances in the growth rate.

The electric field interaction is captured in the $\phi_{n,k}$ term:
\begin{align}
    \phi_{n,k} \equiv \frac{1}{\sqrt{2}} [ E_{k,r} e^{i\phi} J_{n+1}(\xi_j) + &E_{k,l} e^{-i\phi} J_{n-1}(\xi_j)
    \label{eq:phi}
    \\
     + &\frac{v_\parallel}{v_\perp}E_{k,z} J_{n}(\xi_j) ].
    \nonumber
\end{align}
The argument of the Bessel function J is $\xi_j = k_\perp v_\perp / \Omega_j$. Assuming parallel propagating modes with $k_\perp =0$ limits our evaluation to $n=0,\pm 1$ since $J_{n\neq0}(0)=0$. This approximation holds as long as $k_\perp \ll k_\parallel$, as is the case for our calculations.
We further focus on the Alfv\'enic solutions, which limits the calculation to $n= \pm1$ for $\omega_r/k_\parallel \lessgtr 0$ based on the polarization of the modes. \revise{We do not perform a full quasilinear calculation, but only use the quasilinear diffusion operator to determine the regions of phase space responsible for the growth and damping of waves.}

With these caveats, we evaluate $\hat{G}f(k_\parallel d_p,v_\perp/v_A)$ for Interval 2 using the real frequency calculated from the \texttt{ALPS} solutions for the two models and I-I data; these are plotted as color in Fig.~\ref{fig:diffusion1}. We define 
\begin{align}
    \textfrak{G} \equiv \int dv_\perp v_\perp^2 \left| \frac{1}{\sqrt{2}} J_0(k_\perp v_\perp / \Omega_j)\right|^2\hat{G}f_j
\end{align}
and plot this integrated value as a solid teal line in Fig.~\ref{fig:diffusion1}, with surrounding error range in cyan. Over the region in $k_\parallel d_p$ where $\textfrak{G}$ is positive, an instability is present. The error range indicates the bounds for $\pm 10\%$ variation in the retrieved \texttt{ALPS} $\omega_r$ for the Alfv\'en modes to account for any numerical discrepancies in the frequency resolution. The gray shaded regions are areas where \revise{$\gamma_p / \omega_r > 1 / e$}. 

We find in both the bi-Maxwellian models shown in the top two rows of Fig.~\ref{fig:diffusion1} that $\hat{G}f$ has a smooth structure with defined lobes of positive and negative values. For the I-I data in the bottom row, patches of positive and negative regions intermingle and only upon calculation of $\textfrak{G}$ can the final stable or unstable behavior be discerned. The main difference between stable and unstable behavior of the I-I data for the Interval 2 VDF is the presence of a strong negative contribution at small $v_\perp$ (where $v_\perp / v_A < 1$) in the forward Alfv\'en mode. This originates from $G_\perp f$, shown in the top-right panel of Fig.~\ref{fig:diffusionperppar}. Tracing this back to the distribution itself in Fig.~\ref{fig:intVDF}, we see that over the region where Model 2 is unstable (at $v_\perp<$  -34 [km/s]), the I-I data VDF has more variation, particularly at low $v_\perp$. This structure is not captured in either of the models, which show almost symmetric structures in $\hat{G}_\perp$ and $\hat{G}_\parallel$ in the first and second columns of Fig.~\ref{fig:diffusionperppar}. Similarly, in the backward Alfv\'en mode for Interval 2, the models display a smooth structure in both Fig.~\ref{fig:diffusion1} and Fig.~\ref{fig:diffusionperppar} whereas the I-I data structure is far more complex. Both the models and I-I data exhibit unstable behavior for this mode, indicating that the complex VDF structure and gradients does not necessarily suppress an AIC instability and that it may be more coincidence than accurate representation of the I-I data by the models that results in similar unstable behavior for all three distributions.

These features in the diffusion operator support the differences seen in the VDFs in Fig.~\ref{fig:intVDF}, indicating that both the simple contour structure of $f_p$ and the underlying velocity gradients in the I-I data VDF are not being accurately captured by the two bi-Maxwellian models and that different regions and structures in $v_\perp$ and $v_\parallel$ space contribute to the resultant stable or unstable behavior of the plasma normal modes. Even in the case of the two bi-Maxwellian models, very slight differences in the density, temperature anisotropy, and drift speed lead to different behavior, as seen in Fig.~\ref{fig:diffusionperppar}. Slight differences in the relative values of the perpendicular and parallel contributions lead to an overall net negative $\gamma_p$ value, and thus a stable forward Alfvén mode for Model 1, but an unstable mode for Model 2.

\section{Discussion and Conclusions}
\label{sec:discussion}

We utilize \texttt{ALPS} to solve the hot-plasma dispersion relation for inverted-interpolated solar wind proton VDFs based on two intervals observed by the Wind spacecraft and compared these to the dispersion relation for core-and-beam bi-Maxwellian models of these VDFs. These intervals are chosen with $\beta_{\parallel,p}$ and $T_{\perp,p} > T_{\parallel,p}$ such that linear theory predicts that the AIC instability will develop under a bi-Maxwellian assumption.
\revisev{We study each interval using three different VDFs. The first VDF, model 1, is the typical core-and-beam bi-Maxwellian fit to the measured instrument response. We next implement a novel method to extract a more realistic VDF from the Wind Faraday cup measurements; we use a Monte Carlo strategy to apply Gaussian perturbations to a discrete version of model 1. This procedure optimizes the chi-square statistic to produce a VDF that more closely matches the known Faraday cup response. This VDF, which we call the I-I data, better represents the underlying distribution. Our model 2 VDF is constructed from a core-and-beam bi-Maxwellian fit to the I-I data VDF rather than to the instrument response. The more involved procedure used to construct the I-I data yields a distribution with many non-Maxwellian features that reproduces the measured instrument response more accurately than either of the model fits.}

For the two selected intervals, \revisev{a comparison between the model and I-I data VDFs} shows that the bi-Maxwellian models result in fundamentally different plasma behaviors than those calculated for the I-I data they are based on, both by failing to find an instability when one is calculated from the I-I data or by finding significantly different $k_\parallel$ ranges over which the instability operates. 
We show that unstable behavior in Alfv\'en ion-cyclotron waves is sensitively dependent on small VDF structures, particularly in the region where $v_\perp / v_A < 2$.
As bi-Maxwellian models are commonly used to study solar wind proton VDFs, these findings indicate that such simplifications may not accurately represent instabilities in the solar wind, and thus incompletely capture wave-particle energy transfer associated with these VDFs.

We use the quasilinear diffusion operator to demonstrate how the structural differences between the I-I data and model VDFs produce the dissimilar growth rates seen in the \texttt{ALPS} solutions.
Even for modes where both the model and I-I data VDFs yield the same behavior in $\gamma$ (such as the Interval 2 backward Alfv\'en mode where an instability is found in all three cases), calculation of the diffusion operator, $\hat{G} f_j$, for the I-I data reveals a far more complex structure.
These underlying structural differences between the I-I data and model VDFs are apparent even when considering the separate $\hat{G}_\perp$ and $\hat{G}_\parallel$ terms, revealing the failure of these models to truly capture the $f_p$ dependence on either $v_\perp$ or $v_\parallel$ that is found in the I-I data. 

This work demonstrates the importance of including distribution structure when studying the development of microinstabilities. 
The AIC instability threshold, calculated from bi-Maxwellian models of solar wind protons, is frequently exceeded in solar wind observations. Such observations are unexpected as unstable AICs should reduce the temperature anisotropy towards equilibrium.  When we consider a more complex VDF structure (the I-I data in this paper), entirely different stable or unstable AIC behavior is calculated compared to bi-Maxwellian models, potentially explaining this discrepancy. 

\bibliographystyle{apj}
%\bibliography{main.bib}

This work was supported by NASA Grant NNX17AI18G.
K.G.K. is supported by NASA ECIP Grant 80NSSC19K0912. E.L was supported by NSF AGS PRF, Award No. 1949802. 
An allocation of computer time from the UA
Research Computing High Performance Computing at the University
of Arizona is gratefully acknowledged. D.V. was supported by STFC Ernest Rutherford Fellowship ST/P003826/1 and STFC Consolidated Grants ST/S000240/1 and ST/W001004/1. This research was supported by the International Space Science Institute (ISSI) in Bern, through ISSI International Team project \#563 (Ion Kinetic Instabilities in the Solar Wind in Light of Parker Solar Probe and Solar Orbiter Observations) led by L. Ofman and L. Jian. This work was supported by the Open Source Software Sustainability Funding Programme from UCL’s Advanced Research Computing Centre.

\end{document}